# A review on applications of two-dimensional materials in surface enhanced Raman spectroscopy

**Ming Xia**[1, 2]

1   Applied Materials Inc., Santa Clara, California 95054

2   University of California, Los Angeles, Los Angeles, California 90095

*   Correspondence: xiaming@g.ucla.edu

**Abstract:**

Two-dimensional (2D) materials, such as graphene and $MoS_2$, have been attracting wide interest in surface enhancement Raman spectroscopy. This perspective gives an overview of recent developments in 2D materials' application in surface enhanced Raman spectroscopy. This review focuses on the applications of using bare 2D materials and metal/2D material hybrid substrate for Raman enhancement. The Raman enhancing mechanism of 2D materials will also be discussed. The progress covered herein shows great promise for widespread adoption of 2D materials in SERS application.

**1. Introduction**

Raman is a spectroscopy technique that can provide characteristic spectral information of anlaytes. Due to its capability of providing fingerprints of molecule vibration, Raman spectroscopy has a wide variety of applications in chemistry, biology and medicine[1]. However, the yield of Raman scattering is very low leading to weak Raman signals in most cases. Surface enhanced Raman spectroscopy (SERS) makes up this deficiency *via* plasmon resonance. It is capable of ultra-sensitive detection (single molecule detection) and allows for label-free detection with high degree of specificity[2-4]. Molecules absorbed at nanostructured metallic surface experience a large amplification of electromagnetic field due to local surface plasmon resonance leading to orders of magnitude increase in Raman yield and greatly enhanced Raman signal. To achieve high SERS enhancement factors, many efforts have been devoted to develop various metallic (mainly

Au and Ag) nanostructures to enhance the local electromagnetic field[5-9]. Two-dimensional (2D) materials, such as graphene and molybdenum disulfide ($MoS_2$), have unique electronic and optical properties, and thus attract wide interests in their potential applications in electronic devices, sensors, and energy generation [10-15]. In addition to the above mentioned applications, graphene and other 2D materials have also been explored to enhance Raman signals[10, 16-19]. Since the discovery of graphene's Raman enhancement capability[20], extensive researches have been done to reveal the enhancing mechanism of two-dimensional materials, as well as their application in Raman enhancement either with bare 2D material substrate or with hybrid metal/2D material substrate[21-24]. This review will discuss the Raman enhancement mechanism of 2D materials and their applications in SERS.

**2. Raman enhancement mechanism of two-dimensional materials**

Raman enhancement of two dimensional materials is due to chemical mechanism[10, 20, 25, 26], which is different from electromagnetic (EM) enhancement mechanism of most metallic SERS substrates. Chemical enhancement mechanism of SERS has a long controversial history and is still under debate about its nature until recently. Chemical enhancement factor on metallic surface is usually low (~10) compared with EM enhancement factor, which can be as high as $10^6$ to $10^{11}$ [2, 27-29]. In a broad perspective, chemical enhancement can be considered as modification of the Raman polarizability tensor of molecule upon its adsorption, which in turn enhances or quenches Raman signals of vibrational modes[30, 31]. Before the discovery of Raman enhancement of 2D materials, investigation the chemical mechanism is hindered by its low enhancement factor and metal−adsorbate interaction. The fact that 2D materials have no dangling bonds in vertical direction with their atomically flat surface makes them superior platforms to investigate chemical enhancement mechanism. This section will focus on the Raman enhancement mechanism of graphene, hexagonal born nitride (h-BN) and $MoS_2$.

Graphene is the first 2D material that was explored to enhance Raman signals of molecules.[20] Study of copper phthalocyanine (CuPc) /Graphene system reveals that Raman enhancement of pristine graphene is due to the ground state charge transfer mechanism[25]. In ground state charge transfer, adsorbates or analyte molecules do not form chemical bond with SERS substrate necessarily. Graphene is chemically inert and the charge transfer between molecules and graphene causes change in analytes' electronic distribution. Ground-state charge transfer can easily happen between graphene and molecules absorbed on its surface because of graphene's two unique features. One is that the π electrons are abundant on the surface of graphene. The other is that the energy band of graphene is continuous. Figure 1 shows the proposed the ground state charge-transfer process in the graphene enhanced Raman system. In normal Raman scattered process, molecule absorbs photon energy and electrons are excited to a higher-energy level. The electrons then relaxes down the vibrational sub-structure and Raman scattered photons are emitted. The graphene electrons involvement in the Raman scattered process can enhance the electron−phonon coupling, and thus induces the enhancement of the Raman signals. When a vibrational mode involves the lone pair or π electrons, which has stronger coupling with graphene[17, 32], the vibrational mode has highest Raman enhancement. For example, 1530 cm$^{-1}$ mode of CuPc molecules, which represents symmetric stretching of isoindole groups, has highest enhancement than other vibrational modes of CuPc.[32]

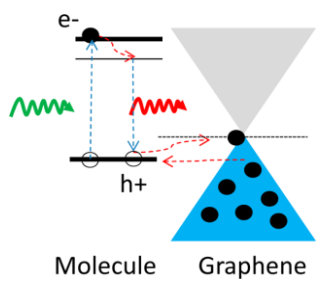

Figure1. Schematic of the Raman scattered process of graphene enhanced Raman spectroscopy.[25]

To gain more insights of the relationship between distance of graphene-analyte and Raman signal, Lin et al.[26] investigated the Raman enhancement of mono- and multilayer ordered aggregates of protoporphyrin IX (PPP) on graphene. The results indicate that most Raman enhancement is from the molecules directly contacting with graphene, which directly proves the "first layer effect" in chemical enhancement mechanism. Functional groups close to graphene layer will be enhanced more than the group further away from the graphene, which again indicates the chemical-enhanced mechanism in this system. As seen in Figure 2, Raman peak intensity of the vinyl group or the porphyrin ring in PPP molecules will be enhanced more when PPP is on bottom position. This is because vinyl group or the porphyrin ring in that position is closer to graphene and the charge transfer is easier to happen. In addition, graphene G-band shifts to 1588 cm$^{−1}$ from 1596 cm$^{−1}$ due to charge transfer between PPP and graphene, revealing the involvement of graphene electrons in the Raman scattered process of analytes.[26]

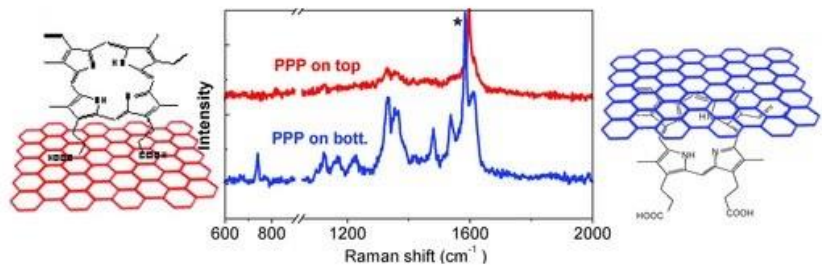

Figure 2. Raman spectra of PPP on the top (red line) or the bottom (blue line) of graphene. The peaks labeled with " ∗ " are the G-band from graphene.[26]

Unlike non-polar and highly conductive graphene, h-BN is highly polar and insulating with a large band gap of 5.9 eV.[33] CuPc molecules are found to be enhanced by h-BN substrates. One proposed Raman enhancement mechanism of h-BN is the interface dipole interaction with analyte molecules, which causes symmetry-related perturbation in the CuPc molecule.[10] In addition, the Raman enhancement factor does not depend on the h-BN layer thickness, because the distribution of the intensity is uniform no matter how thick the h-BN flake is. This result suggests that h-BN is a superior substrate regarding uniformity, when compared to graphene. Atomic layer thin $MoS_2$ is semiconductor and has a polar bond. For $MoS_2$, both the charge transfer and interface dipole interaction are much weaker compared with graphene and h-BN respectively. The Raman enhancement of $MoS_2$ is not as obvious as that of graphene and h-BN, as shown in Figure 3.

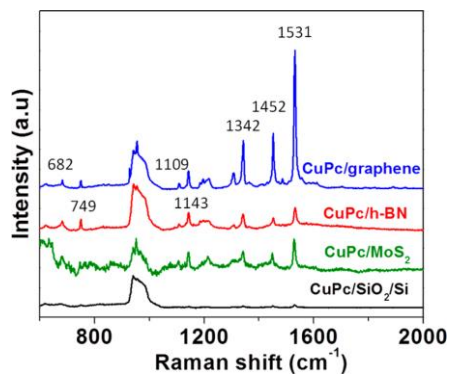

Figure 3. Raman spectra of the CuPc molecule on the blank $SiO_2$/Si substrate, on graphene, on h-BN, and on $MoS_2$ substrates. The numbers marked on the peaks are the peak frequencies of the Raman signals from the CuPc molecule.[10]

**3. Two-dimensional materials alone used as SERS substrate**

Compared with nanostructured metallic SRES substrates, 2D materials can specifically enhance certain vibration modes of molecules due to their chemical enhancement nature. In addition, 2D materials have atomically flat surface and uniform SERS enhancement factors across the whole surface, which benefits the reproducibility of SERS analysis. This section will discuss the application of bare 2D materials in SERS.

Pristine 2D materials have been widely explored to detect molecules. Pristine graphene is the first two-dimensional material used for Raman enhancement. Graphene is found to be capable of enhancing the Raman intensity of various molecules, including phthalocyanine (Pc), rhodamine 6G (R6G), PPP, and crystal violet (CV). Graphene Raman enhancement factors are different for different vibration modes, ranging from 2 to 63.[10, 20] Vibration modes involving the lone pair or π electrons usually have the strongest enhancement because of their stronger coupling with graphene. For instance, 1531 cm$^{-1}$ peak (ring C−C stretch) of CuPc molecules is enhanced more by graphene than 749 cm$^{-1}$ peak (in plane ring symmetric N−M stretch).[10] Graphene substrate is also capable of suppressing fluorescence (FL) background in resonance Raman spectroscopy. Xie et al.[34] demonstrated that graphene could be used as a substrate to suppress FL background by ~$10^3$ times, which can be used to measure resonance Raman spectroscopy from fluorescent molecules such as R6G and PPP.

Besides pristine 2D materials, surface treated and functionalized 2D materials are also explored to improve SERS performance. It is demonstrated that functional group could change the doping level of graphene, and thus modify the SERS signal intensity of R6G molecules.[35] Václav et al [35] demonstrated that fluorinated and 4-nitrophenyl functionalized graphene substrates could provide even higher enhancements for R6G molecules than pristine graphene substrate. UV/Ozone-oxidized Graphene has structural disorder and defects on the graphene surface and resulted in a large chemical mechanism-based signal enhancement.[36] This oxidation along p-doping can result in very high SERS signal. Enhancement factors of UV/Ozone-oxidized graphene could reach ~$10^4$ for molecules like rhodamine B (RhB), R6G, and CV. These functionalization and surface treatment will modify the Fermi level of graphene, and thus affect Raman scattering intensity of molecules on graphene through charge transfer resonance conditions.[37] Surface treated $MoS_2$ shows enhanced Raman intensities of R6G molecules by one order of magnitude compared with pristine $MoS_2$ flakes.[38] The introduction of defects in T-$MoS_2$ samples changes the local surface properties of $MoS_2$ nanoflakes, such as creating the local dipoles, which give rise to the enhancement of Raman signals of R6G molecules and adsorbance of the oxygen in ambient air to dope holes in $MoS_2$, resulting in the enhanced charge transfer effect between R6G and $MoS_2$.

In addition to single type of 2D materials, 2D heterostructure is also explored as a platform for surface-enhanced Raman scattering. Yan et al[39] stacked $WSe_2$ monolayer and graphene together to form a heterostructure for Raman enhancement of CuPc molecules. The results show that the intensity of the Raman scattering on the surface of Graphene/$WSe_2$ (G/W) heterostructure is much stronger compared with isolated layers. The enhanced Raman scattering of the CuPc molecule on the surface of the heterostructure is due to the increasing of the charge transfer at the interface of heterostructure.

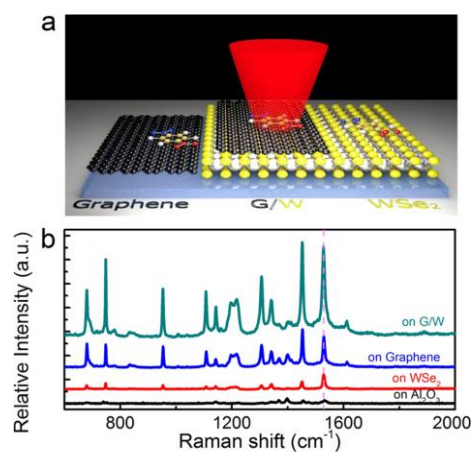

Figure 4. (a) Schematic illustration of the measurement procedure and prepared samples including graphene, $WSe_2$, and G/W heterostructure. (b) Raman spectra of the CuPc molecule on the substrates of the $Al_2O_3$ wafer (black solid line), $WSe_2$ (red solid line), graphene (blue solid line), and G/W (cyan solid line).[39]

**4. Two dimensional materials combined with metallic nanostructures**

Traditional SERS analysis relies on metallic nanostructures that can generate strong local EA field. When combining two dimensional materials with metallic structure, the hybrid SERS substrate can provide even higher SERS EF due to the synergic effect of EM and chemical enhancement. 2D materials, like graphene, could offer chemically inert and biocompatible surface[40, 41], which is favorable in bio-detection. With 2D material as a shielding layer on metallic nanostructure, metal SERS platform such as Ag could be protected from oxidation and have longer shelf life, which can improve the stability and repeatability of the SERS analysis. The following discussion will mainly focus on the incorporation of graphene with various metallic SERS substrates.

Graphene/Au nanopyramid hybrid SERS platform has been explored to detect analytes like R6G and lysozyme, and shows single-molecule detection capability.[17] Even for molecules with small Raman cross-section, like dopamine and serotonin, graphene/Au hybrid platform can still achieve detection limit of $10^{-9}$ M in simulated body fluid.[42] With graphene/Au nanopyramid hybrid SERS substrates, anlaytes peak hot spots and graphene peak hot spots actually coincide as seen from the Raman intensity mapping of analytes peak with that of the graphene G peak (Figure 5). These results indicate that the intrinsic Raman signal of 2D materials in 2D materials/metal hybrid SERS platform could be used as a marker of hotspots. Although FDTD simulation could simulate the EM field of metallic nanostructure and predict the position of hot spots, the actual hot spots of patterned nanostructure could change due to the imperfect nano-fabrication. Graphene and $MoS_2$ have been proven to be capable of overlapping on Au nanostructures and generating strong Raman signals of graphene and $MoS_2$.[43] With 2D materials' intrinsic Raman peak intensity as a SERS EF factor marker, the hot spots of the 2D materials/metal hybrid SERS platform could be located in advance and speed up the later detection of target molecules. This unique feature of hybrid platform offers an advantage for molecule detection in ultra-low concentration. At extreme dilution, the spatial coincidence of an analyte molecule with the hot spots of the hybrid platform will be the key to detect molecular signal. For 2D materials used in hybrid SERS platform with patterned metallic SERS nanostructures, graphene is the ideal choice because graphene only has a few intrinsic Raman peaks and large-area high quality graphene is easily achievable. Another benefit of graphene is that its intrinsic peak intensity could be used to quantify the analytes.[19, 42]

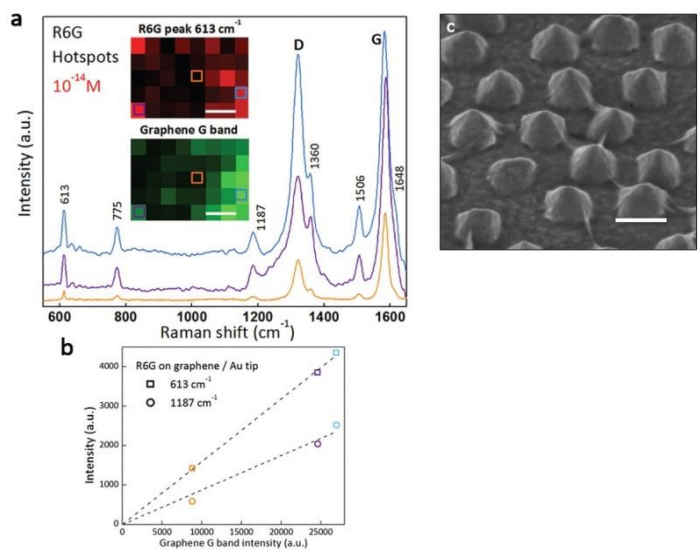

Figure 5. (a) SERS spectra for $10^{-14}$ M R6G on graphene hybrid system taken from three spots from the inset Raman mappings in the same color squares. The inset of panel (a) is composed of Raman intensity mapping of R6G peak at 613 cm$^{-1}$ (red) and Raman intensity mapping of graphene G band (green), scale bar, 2 μm. (b) Raman intensities of R6G peaks at 613 cm$^{-1}$ and 1187 cm$^{-1}$ separately as a function of graphene G band from the three spectra shown in panel (a). (c) SEM image of graphene/Au nanopyramid hybrid SERS platform, scale bar, 200 nm.[42]

The fact that graphene can protect metal to be oxidized has been used in SERS substrate development.[44,45]. Ag nano-structure is known to have excellent SERS performance, but it has one major weakness that is easily to be oxidized in ambient environment. The degradation of Ag will lower the SERS performance and cause uncertainty of analysis. When single layer graphene combines with Ag nanostructure, this hybrid SERS platform could not only provide good SERS performance but also provide excellent stability in a harsh environment (sulfur) and at high temperatures ( 300 °C).[46] Liu at al[45] combined graphene with silver SERS substrates and demonstrated that with the shielding of graphene, the hybrid SERS-Active Substrate could achieve large-area uniformity and long-term stability. As seen in Figure 6, Ag SERS substrates with graphene protection still has high SERS enhacement performance after 6 days exposure to ambient environment, while bare Ag SERS substrate quickly lose its SERS capability by 75% even after one day exposure. The graphene/Ag hybrid SERS platform will benefit the Raman analysis by providing longer shelf life of SERS substrate and improving SERS EF due to chemical enhancement of graphene.

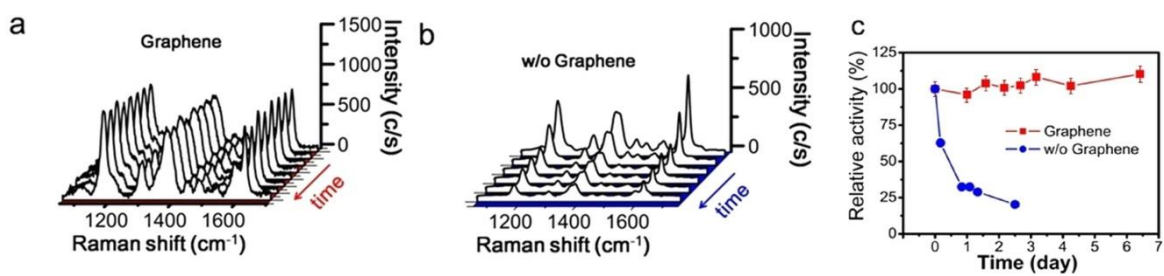

Figure 6. SERS spectra of CV from the SERS substrate with (a) and without (b) graphene protection at different time points. (c) The variation of the SERS intensities at 1167 cm$^{-1}$ versus the time of aerobic exposure (normalized the SERS intensities before the exposure).[45]

Graphene could also be used as a carrier to support metal nanoparticles as SERS substrates. Ag NP/graphene or AuNP/graphene hybrid platform has been prepared and used as SERS platform, showing superior enhancement performance for various molecules.[47-52] With graphene as Ag NP carrier, nanoparticles will uniformly distribute on graphene surface with less aggregation. Therefore, AgNP/graphene provides more hot spots and improved SERS EF. Employing the similar idea, Au NP or Ag NP can also be fabricated on $MoS_2$ sheets to create hot spots and provide enhanced SERS performance.[53-55] Liang et al.[55] has reported a 3D $MoS_2$‐NS@Ag‐NPs nanostructure which can detect trace thiram in apple juice and local lake water with a detection limit as low as 42 nM.

Another advantage to combine graphene and metallic nanostructures it that graphene can help protect molecules from photo-induced damage, such as photobleaching.[16, 56-58] The photobleaching (or photodegradation) of the Raman anlaytes induced by the laser is a well-known phenomenon in normal SERS experiments, especially for dye molecules. When combining graphene with metallic nanostructure, the hybrid SERS platform is more stable against photo-induced damage, but with a comparable enhancement factor. Zhao et al.[58] demonstrated that graphene underneath the organic molecules inhibited the substrate-induced fluctuations; and graphene on top of the organic molecules encapsulated and isolated them from ambient oxygen, greatly enhancing the photo-stability. Liu at al.[56] fabricated graphene-encapsulated metal nanoparticles for molecule detection, and found out that AuNP/graphene hybrid substrate could significantly suppress photobleaching and fluorescence of phthalocyanine and R6G molecules. For instance, within the 160 s measurement period, the 1534 cm$^{-1}$ peak intensity of CoPc molecules decreases dramatically for Au NPs, while the same peak intensity almost keep constant for Au@Graphene, as shown in Figure 7. With graphene as a shielding layer, metal/graphene SERS substrates will provide long-term stability. Du et al[57] discovered the similar phenomenon of graphene protection by simply coating graphene films on Au nanoparticles.

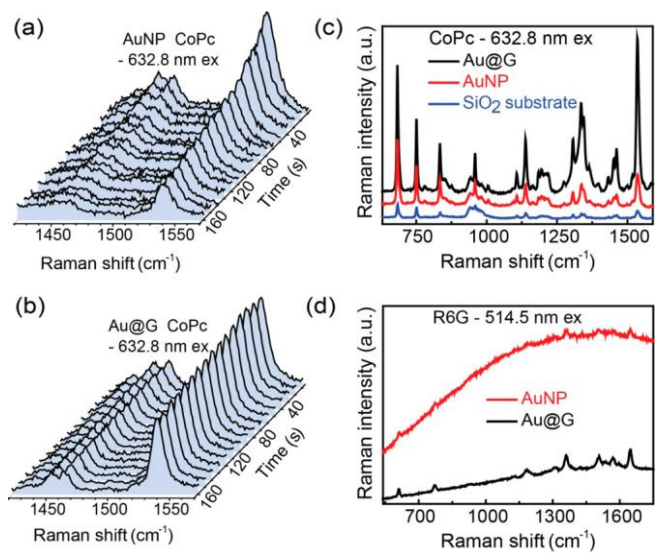

Figure 7. Stability of SERS signals of monolayer CoPc LB films on (a) Au and (b) Au@G. SERS signals of a (c) CoPc film and (d) 3 Å R6G deposited via vacuum evaporation on Au@G and Au NPs.[56]

The benefits of adding 2D materials on metallic nanostructures lay on the following folds. First, 2D materials help further increase SERS enhancement factors due to its chemical enhancement. Although the chemical enhancement factor of 2D materials is not as high as metallic nanostructure, several tens' times of Raman signal enhancement could be essential when detecting molecules at single molecular level. Several times enhancement determines whether the Raman peaks can be seen or not. Second, the atomic thin film of 2D materials will not affect the local EM field of metallic nanostructure and can help map out the hot spots of metallic nanostructure. For instance, monolayer graphene has only 2.3% absorption of the incident laser, and its plasmon resonance frequency is the tetra Hz regime. Therefore, it has little effect on the EM field of underneath metallic SERS substrates. Third, a Raman mapping of graphene G or 2D peak over the hybrid SERS substrates could give the precise position of hot spots. With the information of hot spots location, Raman detection of target analytes could be done only at these hot spots region, and thus improve the detection efficiency especially at ultra-low concentration. Finally, adding graphene as a shielding layer offers chemically inert surface and helps to reduce the fluctuation of SERS signal caused by degradation of the metallic nano-structures, photocarbonization, photobleaching or metal-catalyzed site reactions.

## 5. Conclusions

In summary, 2D materials are unique SERS substrates, whose chemical enhancement nature differentiates them from the majority of metallic SERS substrates. SERS EF of pristine 2D materials is not comparable with that of metallic substrates. However, 2D materials have other advantages over bare metallic SERS substrates, such as chemically inert surface and uniform SERS enhancement. Adding extra atomic layer of 2D materials on surface of metallic SERS structure will help protect SERS metal form oxidation and protect molecules from photo-induced damage with even higher SERS EF due to synergetic effect of chemical and EM enhancement. Combination of 2D materials with metallic SRES structure offers a promising hybrid SERS platform, which can improve the long-term stability and repeatability of SERS analysis.